
\headline={\ifnum\pageno=1\firstheadline\else
\ifodd\pageno\rightheadline \else\leftheadline\fi\fi}
\def\firstheadline{\hfil}
\def\rightheadline{\hfil}
\def\leftheadline{\hfil}
\footline={\ifnum\pageno=1\firstfootline\else\otherfootline\fi}
\def\firstfootline{\rm\hss\folio\hss}
\def\otherfootline{\hfil}
\font\tenbf=cmbx10
\font\tenrm=cmr10
\font\tenit=cmti10
\font\elevenbf=cmbx10 scaled\magstep 1
\font\elevenrm=cmr10 scaled\magstep 1
 1

\line{\hfil }

\vglue 1cm
\hsize=6.0truein
\vsize=8.5truein
\parindent=3pc
\baselineskip=10pt

\noindent
gr-qc/9505011

\noindent
{\it To be published in the Proceedings of the International
Conference on Non Accelerator Particle Physics, Jan. 2-9, 1994,
edited by R. Cowsik (World Scientific, 1995).}
\vskip .5cm

\centerline{\tenbf REALITY AND GEOMETRY OF STATES AND
OBSERVABLES}
\vglue 12pt
\centerline{\tenbf IN
QUANTUM THEORY}
\vglue 5pt
\vglue 1.0cm
\centerline{\tenrm J. ANANDAN }
\baselineskip=13pt
\centerline{\tenit Department of Physics and Astronomy,
University of South Carolina}
\baselineskip=12pt
\centerline{\tenit  Columbia, SC 29208, USA}
\centerline{and}
\centerline{\tenit Department of Theoretical Physics and
Sub-Faculty of Philosophy}
\centerline{\tenit University of Oxford, Oxford, U.K.}
\centerline{E-mail: jeeva@thphys.ox.ac.uk}
\vglue 0.8cm
\centerline{\tenrm ABSTRACT}
\vglue 0.3cm
{\rightskip=3pc
 \leftskip=3pc
 \tenrm\baselineskip=12pt
 \noindent
The determination of the quantum state of a single system by
protective observation is used to justify operationally a
formulation of quantum theory on the quantum state space
(projective Hilbert space) $\cal P$. Protective observation is
extended to a more
general
quantum theory in which  the Schrodinger evolution is
generalized so that it
preserves the symplectic structure but not necessarily the metric
in $\cal P$. The relevance of this more general evolution to
the
apparant collapse of the state
vector during the usual measurement, and its possible connection
to
gravity is suggested. Some criticisms of protective observation are
answered. A comparison is made between the determination of
quantum states using the geometry of $\cal P$ by protective
measurements, via a reconstruction theorem,
and
the determination of space-time points by means of
the space-time geometry, via Einstein's
hole argument. It is argued that a protective measurement
may not determine a time average.

\vglue 0.8cm }
\line{\elevenbf 0. INTRODUCTION \hfil}
\bigskip
\elevenrm

Quantum mechanics was formulated using the beautiful
geometry of Hilbert space [1]. However, much of this mathematical
formalism is deemed unobservable. First, a state vector obtained
by
multiplying a given state vector by a complex number is
physically
indistinguishable from the given state vector. Hence, this complex
number is unobservable. Second, if a particle such as an electron
is in
a state described by a wave function extended over a macroscopic
region in space, and a measurement of the usual kind is made on
the
particle, e.g. let it interact with a photographic plate or cloud
chamber, only a localized spot or trajectory is seen. It appears
therefore that  for some mysterious reason the wave function
collapses to a localized wave packet. Such localized packets, which
give rise to the events that constitute the world we observe, alone
are
observable by the usual measurements performed on a {\it single}
particle.

A third aspect, which is related to the second, is that the wave
function collapses to an eigenstate of the obeservable, represented
by a Hermitian operator in the Hilbert space, in a usual
measurement. Therefore, only these eigenstates are regarded as
``real" in this experiment, because a state $|\psi\rangle$ which is
not
an eigenstate cannot be observed by this measurement performed
on
a single system. Only the probablility of collapse into one of the
eigenstates is predicted. The probability of transition to the state
$|\phi\rangle$ is $|\langle \psi|\phi\rangle|^2$. This is observed
and
can only be observed by performing the measurement on an {\it
ensemble} of identical systems.  The expectation value of the
observable for the state $|\psi\rangle$ is traditionally interpreted
as
a statistical
average of the possible outcomes of the measurement for this
ensemble. Moreover, the observation of the transition probability
determines the absolute value of the complex number $\langle
\psi|\phi\rangle$ but not its phase.

The first aspect of the present formulation of quantum theory
implies that a physical state, or simply a state, must be
represented
by a ray, i. e. a one dimensional subspace, of the Hilbert space
$\cal
H$. The set of physical states, i. e.  the set of rays of the Hilbert
space,
is called the projective Hilbert space and denoted by $\cal P$.

Recently, it was shown that the physical state of a {\it single}
particle
may be observed by performing measurements of a new type
called protective measurement  [2,3], which determines the
expectation value of an observable in the given state instead of
collapsing the state to one of the eigenstates of the observable as
in the usual measurement. This manner of observing the wave
function of a single system
without changing it appreciably will be called a protective
observation. It may
be carried out by having
the system as a non degenerate eigenstate of the Hamiltonian and
performing the measurements adiabatically. It has been extended
to
non degenerate eigenstates and to many particle systems [4]. This
gets around the problem of collapse of the wave function in the
usual
measurement, which is the second aspect mentioned above. Also,
it
gives a new realistic meaning to the points in $\cal P$, because it
is
no longer necessary to use an ensemble of identical systems in a
given state in order to observe this state statistically as
necessitated by the usual measurement. Instead, the state of a
single system is observed.

As mentioned, a protective measurement determines the
expectation value of the
observable in the state being measured. It therefore extends the
reality given to the eigenvalues of the observable being measured
in
the usual formulation of quantum theory to all expectation values
of
the observable.  Hence, it overcomes the limitation in the
observable
quantities in the usual formulation mentioned as the third aspect
above. Also it suggests that an observable may be represented by
its
expectation value, which is a real valued function of $\cal P$,
because it is this function which is determined by experiment.

Here I shall briefly review the protective
measurement and describe and a feasible experimental realization
of it in section 1. A reconstruction theorem which assures that the
expectation values contain all the kinematical and geometrical
information of quantum mechanics is stated. Quantum mechanics
is formulated in $\cal P$, in section 2, using the natural symplectic
structure defined by the inner product in the underlying Hilbert
space. This leads to a natural generalization of quantum theory in
section 3. In section 4, some criticisms of the use of protective
measurement so far are shown to be not valid. It is shown, in
section 5, that the determination of states by their relations to
other states by means of protective measurements via the
reconstruction theorem, is analogous to the determination of
space-time points from a solution of gravitational field equations
which are generally covariant. In section 6, I argue against the
possibility that a protective measurement measures a time
average
of an observable of a single system, which would have enabled
one to give statistical meaning to this type of measurement.
\vglue 0.8cm
\line{\elevenbf 1. PROTECTIVE OBSERVATION \hfil}
\bigskip
\elevenrm

The wave function undergoes two types of evolution in quantum
theory, according  to the usual Copenhagen interpretation. 1) It
undergoes a continuous, linear, reversible, deterministic evolution
according to Schrodinger's equation when no ``measurement" is
being
made.  2) It undergoes an apparantly discontinuous, non linear,
irreversible, indeterministic evolution during a ``measurement".
i.e.
the evolution (2) is the negation of (1) in every way. A
``measurement" at present is defined only in a very vague sense
as
the quantum system coming into contact with a macroscopic
apparatus, which is not precisely defined. This causes a ``collapse"
of
the wave function, which is necessarily preceded by the combined
system being in an entangled state as a result of the evolution (1).
The outcome of this collapse can
only
be predicted probabilistically. This leads to the well  known
statistical interpretation of the wave function.

This state of affairs is unsatisfactory. But on the other hand
measurement plays a very important role in physics, not only in
obtaining information about the particular system being observed
but also in providing operational meaning to the concepts of
physics.
This makes it imperative to look for a measurement that uses only
the evolution (1), and avoids the collapse (2). Protective
measurement, which will be described now, is such a
measurement.

A protective measurement determines the expectation value of an
observable for a wave function
while it is prevented from collapse because of another interaction
it
undergoes at the same time. For example, the system may be
protected by being prepared in an eigenstate of a Hamiltonian,
and
the measurement is made adiabatically. Then no entanglement
takes
place between the system and the apparatus. Therefore no
collapse
takes place either.

Some examples of protective observations were given in ref. 3. A
protective observation may be made, in principle, on the
polarization
state of a
photon, analogous to the protective observation of a spin 1/2
particle [3] as follows [5]: Send a circularly polarized photon
through a
crystal with large optical activity. This crystal has different
refractive
indices for
the two orthogonal circularly polarized states. Make the crystal
also
have an
elliptical birefringence which varies in space. This would have
different
refractive indices for the two corresponding orthogonal elliptically
polarized states.
Since the refractive indices vary in space, the two elliptically
polarized
states ordinarily get separated in space analogous to the usual
Stern-Gerlach experiment. But because of the large optical activity,
the circularly polarized state is protected, and this
separation therefore does not take place. This is analogous to
protecting the spin state of the neutron by a large magnetic field
in the direction of the spin. But the trajectory of the
center of mass of the photon is curved, because of the spatially
varying elliptical birefingence, in a manner that is determined by
its
polarization state. By oberving this trajectory, it would be possible
to determine the polarization state of a
single photon analogous to the observation of the spin state
of a single neutron in the protective Stern-Gerlach
experiment [3,4].

 The expectation value of an observable  $A$ will be denoted by
$\langle A\rangle$, and its value for a given wave function $\psi$
is
defined to be
$$\langle A\rangle(\psi) \equiv {\langle \psi|A|\psi \rangle\over
\langle \psi|\psi
\rangle}. \eqno(1.1)$$
It is clear that $\langle A\rangle$ is constant on any ray (one
dimensional subspace of the Hilbert space), and therefore it may
be
regarded as a function of the set of rays, called the projective
Hilbert
space $\cal P$.
By protectively measuring the expectation values of a sufficient
number of such observables, the ray containing
the wave function of a
single system can be determined.

The term `expectation value' is an unfortunate choice of
terminology for (1.1) because it conveys the meaning that it is a
statistical average. This was justified prior to the discovery of
protective observation because the only way of giving physical
meaning then to (1.1) was by making measurements on an
ensemble of identical systems in the given state. But since
protective measurement enables (1.1) to be determined directly
for a single system, I shall refer to any particular value of (1.1) as
an
observed value. Also, the function of $\cal P$ defined by (1.1) will
be called an observable.

It is a remarkable fact that the observables, whose values for
physical states may be obtained in principle by an experimentalist
via protective measurements without assuming any underlying
Hilbert space structure, completely determine the Hilbert space
underlying $\cal P$ and its geometry. Moreover, each observable
then is given by (1.1) for a suitable Hermitian operator $A$ that
acts on this Hilbert space. This follows from the following theorem
[4]: From the observables given as functions on $\cal P$, regarded
as an abstract set, the vector
space structure and the inner product in
the Hilbert space can be reconstructed. They
are then respectively unique up to
isomorphism and multiplication by a real positive constant.
Also, after the reconstruction of the Hilbert space $\cal H$, each
observable uniquely determines the corresponding Hermitian
operator in $\cal H$ so that they satisfy (1.1) for each state.

Since the observed values can be measured non
statistically for a {\it single system} using protective
measurements, the above reconstruction theorem provides
the states and observables with a
new ontological meaning. This will be discussed in more detail in
section 5. Historically, there has been a close affinity between
such an operational approach and the geometrical approach to
physics. For example, in the creation of special relativity, the
realization by Einstein of the problem of simultaneity and that the
$t$ and $t'$ in the Lorentz transformation which are measured by
clocks in equivalent ways are on the same footing, led to
Minkowski geometry. The probing of gravity by classical particles
led via the equivalence principle to curved space-time geometry.
So, there appears to be a metatheoretical principle that ensures
that the operational approach naturally leads to a geometrical
description.
It should be of no surprise therefore that protective
measurements, which determine the observed values as functions
of $\cal P$ for a single
system should lead naturally to a corresponding geometric
reformulation of quantum mechanics on $\cal P$. This will be
discussed in the
next section.
\vglue 0.6cm
\line{\elevenbf 2. QUANTUM MECHANICS IN THE QUANTUM
STATE
SPACE \hfil}
\vglue 0.4cm

The vector space structure of the underlying Hilbert space $\cal
H$
endows the quantum state space $\cal P$
with a projective geometry. It was for this reason that $\cal P$
was called the
projective Hilbert space [6]. The projective geometry of $\cal P$
may be defined, in the sense of Klein's Erlanger program, as the
set of properties that are invariant under the group of non
singular linear transformations acting on $\cal H$. It may be
viewed roughly
as the geometry obtained from Euclidean geometry by removing
the
concepts of distance, angle and parallelism. For example,
collinearity
is a projective property.

The fundamental physical property of states which gives rise to
quantum theory is the phenomenon of interference. Algebraically,
this reflects the fact that a complex linear combination
$$|\psi> = c_1 |\psi_1> + c_2 |\psi_2> \eqno(2.1)$$
of two state vectors $|\psi_1>$ and $|\psi_2>$ represents a new
possible physical state.
The relation (2.1), being invariant under linear transformations, is
a projective property. The physical states, or rays to which
$|\psi_1>$, $|\psi_2>$ and $|\psi>$ belong to, are said to be three
collinear points in $\cal P$, with respect
to the projective geometry in $\cal P$. The ratio $c_2\over c_1$,
which is {\it not} a projective invariant, may be regarded as the
inhomogeneous coordinate of the physical state of $|\psi>$ in the
coordinate system on this line determined by the physical states
of
$|\psi_1>$ and $|\psi_2>$. Conversely, given any three collinear
points on $\cal P$, each
of the three states may be regarded as an appropriate
superposition
of the other two states. This gives a geometrical meaning to
quantum interference.

Two more important geometrical structures in $\cal P$ are
determined
by the inner product in $\cal H$. These will be introduced from a
physical point of view in the remainder of this section. This inner
product between any
two state vectors has the physical meaning of the probability
amplitude for the transition between the two corresponding state
vectors. Its absolute value squared has a direct physical meaning
as the probability of transition when the usual measurement is
performed, which is given physical meaning by means of
measurements performed in an ensemble of identical systems. But
the reconstruction theorem stated in section 1 shows that the
entire probability amplitude, and not just its absolute value, may
be given a physical meaning associated with a single system by
means of protective measurements.

Another physical meaning for the inner product is provided by
the geometric phase [7,6]. Suppose a state vector undergoes cyclic
evolution, i.e. the initial and final state vectors are the same
except for
a phase. This phase contains a geometric part, the geometric
phase, that depends only on the motion of the state that is given
by an unprarametrized closed curve $\gamma$ in $\cal P$ and
not
on which of the infinite class of possible Hamiltonians actually
evolves the system along this curve. Choose $|\tilde \psi>$
belonging to the points in a neighborhood containing $\gamma$ as
a differentiable normalized
function of this neighborhood. Then the geometric phase
associated with $\gamma$ is [6]
$$\beta = \oint_\gamma i<\tilde \psi|d\tilde \psi>. \eqno(2.2)$$
Now $\exp (i\beta )$ is the holonomy transformation when a state
vector $|\psi (s)>$ is parallel transported along $\gamma$
according to the rule that the inner product between neighboring
states $<\psi (s)|\psi (s+ds)>$ is to first order real and positive [8].
For this
reason it provides a fairly direct measurement of the inner
product, including its phase. But $\beta$ is also the symplectic
area of any surface
spanned by
$\gamma$ with respect to a natural symplectic structure in $\cal
P$ defined by the inner product in $\cal H$ [9,10,11]. So, again,
seeking
a direct physical meaning for mathematical structures in $\cal H$
has led us to a geometrical structure in $\cal P$ .

I shall now express this symplectic structure [12] in a language
familiar to physicists [13]. Let $\tilde\psi_j$ be the components of
$|\tilde \psi>$ with respect to an orthonormal basis. Then
$\tilde\psi_j$ may be regarded as homogeneous coordinates in
$\cal P$. Define $Q^j =\tilde\psi_j$ and $P_j =i\tilde\psi_j^*$,
where
the $*$ denotes complex conjugation. Then (2.2) may be rewritten
as
$$\beta = \oint_\gamma P_j dQ^j =\int_\Sigma dP_j \wedge dQ^j ,
\eqno(2.3)$$
where $\Sigma$ is a surface spanned by $\gamma$. Specifying a
symplectic structure in $\cal H$ is equivalent to specifying a
preferred class of canonical coordinates $Q^j$ and momenta $P_j$
using which the experimentally observable $\beta$ may be
expressed in the form (2.3) for any cyclic evolution. The above
choice is a special case for which the coordinates and momenta are
complex. The 2-form $\Omega \equiv dP_j \wedge dQ^j$ is called
the symplectic
2-form which equivalently determines the symplectic structure.
But (2.2) and therefore (2.3) is independent of the choice of
$|\tilde\psi >$. Therefore, we have a symplectic structure in $\cal
P$
because the canonical invariant (2.3) depends only on $\cal P$. If
$\cal H$ has $N+1$ complex dimension then there are $2N+2$ real
coordinates in $Q^i$
and $P_j$. But the normalization of $|\tilde \psi>$ and the choice
of phase for $|\tilde \psi>$, e.g. $\tilde \psi_1$ may be chosen to
be real, implies that only $2N$ of these coordinates are
independent. This makes $\cal P$, which has $N$ complex
dimensions, a $2N$ real dimensional phase space.

All of quantum mechanics may be formulated entirely on $\cal P$
using the above symplectic structure. Suppose $f$ and $g$ are two
functions on $\cal P$. The Poisson bracket between them is
defined in the usual way:
$$\{ f,g\}={\partial f\over \partial Q^j}{\partial g\over
\partial P_j} - {\partial f\over \partial P_j}{\partial g\over
\partial
Q^j}, \eqno(2.4)$$
with summation over repeated indices. Given two observables
$<A>$ and $<B>$, which are functions of
$\cal P$, formed from the operators $A$ and $B$, it is easily
verified that
$$\{ <A>,<B>\} =<[A,B]> . \eqno(2.5)$$
The Hamiltonian observable is $<H>$ which corresponds to the
Hermitian Hamiltonian operator $H$. It is convenient to choose
$|\psi>$ in the definition of $<H>$ to be normalized. Then
Schrodinger's equation
is equivalent to the Hamilton's equations
$${dP_i\over dt}=-{\partial <H>\over \partial Q^i},{dQ^i\over dt}=
{\partial <H>\over \partial P_i}.\eqno(2.6)$$
For the complex $Q^i $ and $P_j$ chosen above, the two equations
(2.6) are Schrodinger's equation and its complex conjugate.
Alternatively, we
may choose $Q^i =\sqrt{2}{\rm Re}\tilde\psi_j$ and $P_j
=\sqrt{2}{\rm Im}\tilde\psi_j$, which are real, in which case the
two
equations (2.6) are independent and represent the real and
imaginary parts of Schrodinger's equation.

On choosing local
canonical coordinates $X^A = (P_i,Q^j)$ and defining $h=<H>$, (2.6)
may be recast in the more compact form
$${dX^A\over dt} =\Omega ^{AB}{\partial h\over \partial X^B} ,
\eqno(2.6') $$
where $\Omega ^{AB}$ is the inverse of the symplectic 2-form
$\Omega _{AB}$.
The  Poisson bracket (2.4) may now be rewritten as
$$\{ f,g\}=\Omega ^{AB}{\partial f\over \partial X^A}{\partial
g\over
\partial X^B}.\eqno(2.4')$$
Then, it follows from $(2.6')$ that the time evolution of an
arbitrary
observable $a$ is given by
$${da\over dt}=\{ a,h\}.\eqno(2.7)$$
An advantage of formulating quantum theory in terms of (2.7) is
that {\it it abolishes the distinction between the Schrodinger and
Heisenberg pictures}. Because in both pictures the `expectation
values' of the observable operators evolve in the same way, given
by (2.7).

The inner product also gives the metric on $\cal H$
$$d\sigma^2 = \sum_k \{(dQ^k)^2 + (dP_k)^2\} =\delta_{AB}dX^A
dX^B .\eqno(2.8)$$
The requirement $<\tilde\psi|\tilde\psi> = 1$, i.e.
$$\sum_k \{(Q^k)^2 + (P_k)^2\} =1 ,\eqno (2.9)$$
defines the unit sphere $\cal S$ which is a submanifold of $\cal
H$. The set of states $\{e^{i\theta}|\tilde\psi>\}$ of $\cal S$ for a
given $|\tilde\psi>$ is a point of $\cal P$. So, $\cal S$ is a principal
fiber bundle over $\cal P$ with structure group $U(1)$
corresponding to this multiplication by $e^{i\theta}$. A connection
in this bundle may be defined by specifying the horizontal spaces
to be normal to the fibers with respect to the metric (2.8).

If $\Pi$ is the projection map of the bundle $\cal S$ into $\cal P$,
then it is natural to define a metric on $\cal P$ by requiring that
$\Pi_*$ when restricted to each horizontal space is an isometry
with respect to this metric.
This is a Kahler metric on $\cal P$, called
the Fubini-Study metric.
More explicitly, the metric (2.8) when restricted to $\cal S$ is [14]
$$ d\sigma^2 = (d\phi-A)^2 + dS^2 , \eqno(2.10)$$
where $\phi$ is the phase of $\tilde\psi_1$, $A=i<\tilde
\psi|d\tilde \psi>$ and $dS^2$ is the metric on the horizontal
space, which is the same as the Fubini-Study metric on $\cal P$
by the above
construction. A horizontal vector is annhilated by the connection
1-form $A$. By integrating this equation, the geometric phase
(2.2) may be obtained as the holonomy associated with the closed
curve $\gamma$ of $\cal P$.

Suppose $\tilde\psi_k$ are chosen to be the components with
respect to an orthonormal basis of eigenstates of the Hamiltonian
with eigenvalues $\omega_k$. Also, we may choose $\phi =0$.
Then $X^A$ may be regarded as local coordinates on $\cal P$. The
Hamiltonian function on $\cal P$ is
$$h={1\over2}\sum_k \{ (Q^k)^2 + (P_k)^2\}\omega_k.
\eqno(2.11)$$
This is like the Hamiltonian for a set of non interacting harmonic
oscillators.
The Schrodinger time evolution,
generated by a Hermitian Hamiltonian is an isometry in $\cal P$
with respect to the Fubini-Study metric [15]. The evolution of a
given initial
state is a curve in $\cal P$. It is a remarkable fact that the time
integral of the uncertainty of energy along this curve is
independent of which Hamiltonian is used for this evolution.
Therefore, like $\beta$, it must have a geometrical meaning but
now even when the curve is not closed. It is in fact the distance
along the curve measured by the Fubini-Study metric.

To
conclude, the inner product in $\cal H$ gives a symplectic
structure and a Kahler metric in $\cal P$, both of which can in
principle be generalized to obtain physical theories more general
than the present day quantum theory.

\vglue 0.6cm
\line{\elevenbf 3. POSSIBLE GENERALIZATION OF QUANTUM
THEORY \hfil}
\line{\elevenbf AND THE MEASUREMENT PROBLEM \hfil}
\vglue 0.4cm

The above geometric reformulation of quantum theory using the
symplectic structure suggests a natural generalization of quantum
theory [12]. We may require only that
the time evolution preserves the symplectic structure, and need
not preserve the metric. Then the evolution is given by $(2.6')$
with
$h$ being any real differentiable function of $\cal P$ which need
not
be the `expectation value' of a Hermitian Hamiltonian operator. I.e.
$h$ may be more general than the function (2.11).
This
would
allow for time evolutions which are non unitary and non linear in
$\cal H$.

This generalization of the Hamiltonian evolution also leads
naturally
to a generalization of an observable $a$ as any real differentiable
function on
$\cal P$ that need not have a representation of the form (1.1).
Such an observable may in principle be observed by protective
measurements as follows. Couple the system to an observable $q$
of an apparatus, for example its `pointer position', so
that the Hamiltonian of the combined system is
$$h =h_0 + g(t) q a + h_A , \eqno(3.1)$$
where $h_0$ and $h_A$ are the unperturbed Hamiltonians of the
system and the apparatus respectively, and $g(t)$ represents the
turning on and off of the interaction between them during a
time interval, say $[0,T]$.

Suppose, for simplicity, that $h_0=<\psi|H_o|\psi>$ and
$h_A=<\psi|H_A|\psi>$, where $H_0$ and $H_A$ are Hermitian
operators, and $<\psi|\psi>=1$. I.e. $h_0$ and $h_A$ are like the
observables in ordinary quantum theory, but $a$ is assumed to be
a more general observable. Let $|n>$ be the eigenstates of $H_0$
with eigenvalues $\omega_n$. An arbitrary evolution of the
system may be represented by the state vector
$$|\psi(t)>=\sum_j c_j(t)\exp (-i\omega_j t)|j>.\eqno(3.2)$$
I shall choose complex coordinates with respect to the basis $|n>$,
as defined in section 2.
So, the evolution of the system, in terms of these coordinates, is
given by $Q^j(t) = c_j(t)\exp (-i\omega_j t), P_j(t) =
ic_j^*(t)\exp (i\omega_j
t)$. Then, substituting this in the generalized Schrodinger
equation $(2.6')$ yields
$$c_j(T)=q\int_0^T g(t)\exp (i\omega_j t){\partial a\over \partial
P_j}.\eqno(3.3)$$

Suppose now that the interaction Hamiltonian is small compared
to the protection Hamiltonian contained in $h_0$. Then first order
time dependent perturbation theory may be used. This consists in
approximating the $Q^j$ and $P_j$ in the right hand side of (3.3)
by their unperturbed values. If the intial state vector of
the system was $|s>$, then these unperturbed values are $Q^j =
\delta_{js} \exp (-i\omega_s t), P_j = i\delta_{js}\exp (i\omega_s
t)$.
Now if, as in the usual quantum mechanics, $a=<\psi|A|\psi>=-
i\sum_{j,k} P_j<j|A|k>Q^k$ for some Hermitian operator $A$, then
the unperturbed value of ${\partial a\over \partial P_j}$ is $-
i<j|A|s>\exp (-i\omega_s t)$. Owing to the reality of $a$, the same
time dependence may be assumed for the present more general
case of $a$ being an arbitrary real differentiable function of $\cal
P$. Then from (3.3), $c_j(T)$ is proportional to the Fourier
component of $g(t)$ corresponding to the transition frequency
$\omega_j-\omega_s$.

So, if $\omega_s$ is a non degenerate eigenvalue and $g(t)$ is
sufficiently slowly varying then these Fourier components for
$j\ne s$ are negligible. Then the system will remain in the original
state $|s>$, without entanglement with the states of the apparatus.
This entanglement is a prerequisite for the collapse of the state
vector. Therefore, there will be no collapse in this case. But as
pointed out [4], neither the non degneneracy nor the adiabaticity
assumption is necessary to ensure this non entanglement. It is
sufficient if $g(t)$ has neglible Fourier components corresponding
to all the transition frequencies, as shown also by the above
treatment. If this is not satisfied then there would be transition to
other energy eigenstates of $H_0$, which would result in
entanglement.

In the present case in which there is no entanglement, a
protective
measurement of $a$ may be made by measuring the change in the
observable $p$ of the apparatus conjugate to $q$, i.e. $\{p,q\}=-1$.
The equation of motion for $p$ is  given by (2.7) to be
$${dp\over dt}=-g(t) a .\eqno(3.4)$$
Therefore, the change in $p$ is $\Delta p =-g_o a$, where
$g_0\equiv \int_0^T g(t) dt$ is known. Hence, by measuring
$\Delta p$, $a$ may be determined for a single system. So, the
present generalized quantum mechanics permits the
determination of the more general observables for a single
system, in principle. This generalizes protective measurement to
the measurement of these generalized observables.

In particular, the generalized hamiltonian $h$ in the present
generalization of quantum theory may be observed, in principle,
by means of
protective measurements on a single system. This together with
the possibility, in principle, of protectively observing the quantum
state at any given time would justify regarding the state and its
time evolution to be associated with a single system. The question
arises if such more general time evolutions would account for the
collapse of the state vector of individual systems which appears to
occur in a non linear manner.

The possible connection between this question and quantum
gravity may be seen from the following consideration. From a
physical point of view, the arena for the geometry of quantum
theory is $\cal P$ as I have argued in section 2. This is very
different from the arena of space-time for the geometry of
general relativity. But contact is made between the two
geometries when a measurement is made on a particle. Because
then the localization of the state vector, as it collapses, provides an
approximate
event which is a point of the space-time manifold that
incorporates gravity. So the collapse of the wave function provides
a connection between quantum theory and gravity, even at an
energy scale which is far below the scale at which high energy
physicists expect quantum gravitational effects to become
important. This argument suggests that the scheme of Penrose
[16] and others to explain the collapse of the state vector using
quantum gravity is worth examining carefully.

\vglue 0.6cm
\line{\elevenbf 4. REPLY TO CRITICISMS OF PROTECTIVE
OBSERVATION \hfil}
\vglue 0.4cm

As originally formulated [2,3], in a protective observation the
wave
function of a particle is protected by an experimentalist who then
gives it to another observer. This observer is only told that the
wave
function is protected but not what the protecting Hamiltonian is.
She
then determines the wave function by measurements on the
given single particle. This has the novel aspect that the wave
function is determined for a single particle and not an ensemble of
identical particles as in all previous measurements. Nevertheless,
this
has been criticized by several physicists on the following grounds:
(a) The
experimentalist who protects the wave function has at least a
partial
knowledge of the wave function and therefore it cannot be said
that
the subsequent measurement by the observer determines a
completely unknown wave function [17,18]. (b) If the wave
function already comes protected, then the experimentalist
performing the subsequent protective measurement is playing a
passive role in that she is not providing the protection [19].

But the purpose of doing a measurement is to (i)
determine the state of the system and (ii) determine the
manifestation of this state in order to infer its physical meaning.
Protective measuements show the manifestation
of the wave function of a single particle, which has not been done
before. By this I mean that the observed (expectation) values of
not
necessarily
commuting Hermitian operators may be measured by protective
measurements on a single particle. From these measurements of
sufficiently large number of observables the wave function can be
reconstructed for this particle. Using the usual measurements, the
expectation value could be given physical meaning only as a
statistical average of measurements on an ensemble of particles.
So,
protective measurement, unlike the usual measurement, fulfills
the
goal (ii) for a single particle.

Another criticism was that the registration in the final stage of the
protective measurement constitutes a measurement of the usual
type
involving an irreversible amplification and possibly a collapse,
and
therefore the protective measurement does not circumvent the
measurement problem [20]. But the state being protectively
measured
does not become entangled with the apparatus state. Therefore,
the
collapse of the apparatus state does not affect the protected state.
An
ensemble of apparatii may be used to measure the apparatus
state as
in  the usual measurement, but they need to be coupled only to
one
observed system whose state, being protected, does not collapse.

For example, in the usual Stern-Gerlach experiment of a spin 1/2
particle, quantum theory predicts an entangled state which
corresponds to two spots. After the irreversible amplification by
the
detector D, only one spot is observed. Hence the entangled state is
collapsed, in the Copenhagen interpretation, to have agreement
with
observation. The search for a description or explanation of the
collapse is the measurement problem. But in a protective
Stern-Gerlach experiment [3] only one spot appears
whose position, which is between the two spots of the usual
measurement, is determined by the spin state. So, there is no
entanglement. The inclusion of the detector D does not change this
in
any way, because during the resulting irreversible amplification,
the
combined system is still evolving in a non entangled state
corresponding to one spot only. I.e. D has only one possible
``pointer
reading" according to Schrdinger's equation, and therefore there
is
no need to collapse the wave function. The formation of the single
spot takes place exactly in the same way as the formation of
either of
the two possible spots in the usual Stern-Gerlach experiment, but
the
entanglement of the latter two spots is avoided.

The above criticisms (a) and (b) have been completely
overcome by the reconstruction theorem [4], stated in section 1,
that from the observed
values obtained as functions of the abstract set of states $\cal P$,
the Hilbert space together with the inner
product can be constructed. Moreover, the operators whose
observed values were experimentally obtained are then uniquely
determined as Hermitian operators in this Hilbert space. Now the
physical and geometric meaning for the various possible states are
obtained by the observed values and the inner products
between
state vectors. Therefore, the same experimentalist may both
protect
and measure each state without knowing prior to the
measurement
anything about the structure of $\cal P$ or the underlying Hilbert
space, or the observable operators
associated with the apparati used to make the
measurements. Then the above mentioned reconstruction theorem
enables the meaning of the states to be determined in a wholistic
manner. Since the protective measurements may all be done on a
single system, the states and observables may therefore also be
associated with a single system.
\vglue 0.6cm
\line{\elevenbf 5. COMPARISON WITH EINSTEIN'S (W)HOLISTIC
ARGUMENT \hfil}
\vglue 0.4cm

The wholistic manner in which protective measurements
determine
the state of the system, described in section 1, may be
compared with the resolution of a paradox due to Einstein. In
1913,
Einstein argued that if the gravitational field equations were
generally covariant then for the gravitational field inside a hole of
some given distribution of matter, there would be infinite number
of
solutions related by diffeomorphisms [21]. This  is  unlike in the
electromagnetic case in which there is a unique solution for the
electromagnetic field inside a hole surrounded by a given charge
distribution. So, Einstein concluded incorrectly that the field
equations for gravity cannot be generally covariant, which
delayed
the discovery of general relativity by two years.

In 1915, Einstein realized that this argument is not valid for the
following reason: The points of space-time should be determined
operationally as ``{\it spatiotemporal coincidences}''. Such a
specification of points, as an intersection of two world-lines, is
invariant under diffeomorphisms. This implies that a space-time
point may not be given reality as an independent isolated object.
It may however be given reality
wholistically using the geometry determined by the metric. E.g. a
given
point inside the hole may be determined by the distances
along
the geodesics joining this point to identifiable material points on
the boundary of the hole. Then the
infinite number of solutions related by diffeomorphisms
correspond
to the same objective physical geometry.

Similarly, the physical states in quantum
theory, which are points of the projective Hilbert space, should be
determined by their relations to other points through the
geometry of the projective Hilbert space.
The reconstruction theorem stated in section 1 shows how the
geometry of $\cal P$ may be determined by protective
measurements on an individual system. Then each state acquires
physical meaning through its relationship to other states which is
determined by this geometry.

This way of determining the state may be compared with
someone
being shipwrecked in a small island. To determine where this
island
is it would be necessary to know its location with respect to other
lands. Similarly, to know what the state is, one needs to know its
relation with other states. I. e. the structure of the projective
Hilbert space
needs to be determined.

It is better to give a single particle interpretation right from
the very beginning to the Hilbert space by means of the
reconstruction theorem, rather than to first construct the Hilbert
space on the basis of usual measurements which need to be done
on
ensembles of identical systems, and then to show how the state of
a single particle may be protectively observed. Because
in
the latter case, which was originally done, the Hilbert space
machinery used in reconstructing the state used
ensembles
and therefore it could be argued that the reconstructed state
cannot be associated with a single system.
\vglue 0.6cm
\line{\elevenbf 6. DOES PROTECTIVE MEASUREMENT \hfil}
\line{\elevenbf DETERMINE A
TIME AVERAGE? \hfil}
\vglue 0.4cm

Let us examine the possibility that in a
protective measurement of the expectation value of an observable
$A$ for a single system what is obtained is a time average of $A$,
which is equal to the ensemble average. This would be like
following
the motion of  an individual molecule in a classical gas. If we
follow it
long enough, its time averaged properties would be like as if it
takes
all the positions and velocities of the molecules at a given time.
This
is why in classical statistical mechanics the time average equals
the
ensemble average. If a similar situation exists for a protective
measurement, then it would not remove the statistical element of
quantum theory.

It was shown in ref. 3, however, that it is not possible to regard
the outcome of a protective measurement for a particle in a box as
a time average of a classical variable. Consider now whether a
general protective measurement of an observed value may be a
time average of the quantum mechanical state treated
ontologically. Then the system with state vector $|\psi >$ should
pass through
each of the eigenstate vectors $|\psi_i>$ of the observable $A$
spending times at these
states proportional to $|<\psi|\psi_i>|^2$, in order that the time
average
equals the ensemble average. This would require a drastic
modification of quantum theory. In any case, this possibility may
be
precluded by monitoring the state during the measurement
and ensuring that it is in the state corresponding to $|\psi>$ and so
does not pass through the states $|\psi_i>$.

But it may be argued that the observed value is the time average
of possible outcomes of measurement of the state that is treated
epistemologically. This means that the state may be exactly as
predicted by Schrodinger's equation, but that it does not describe
a real process. Its meaning is only that it predicts the probabilities
of possible outcomes of measurement. These pobabilities may be
given physical meaning as the average of measurements over an
ensemble of identical systems having this state at a given time or
as a time average of measurements performed on a single system
that is protected over a long period of time.

This view is not tenable, however, for the following reason. A long
period of time $T$ here means that

$$T>>{\hbar \over \Delta E} , \eqno(6.1)$$
where $\Delta E$ is the smallest of the energy differences
between the protected state and the other energy eigenstates.
But there are an infinite number of Hamiltonians, with infinite
values
for $\Delta E$, which have the protected state as an eigenstate
with the same eigenvalue. The time evolution of the protected
state is the same for all these Hamiltonians. But the time scale
which determines how long the measurement should be, namely
${\hbar \over \Delta E}$, is different for these Hamiltonians. It is
therefore not a property of the state being observed protectively.
Indeed, $T$ may be
made as small as one pleases, in principle, while satisfying (6.1)
by making $\Delta E$
correspondingly large. Hence, the outcome of a protective
measurement cannot be regarded as the time average of a
property of the state.

Also, as mentioned in section 3, adiabaticity is not needed for
protective measurement, provided the interaction with the
apparatus does not cause transition between the original and
neighboring energy eigenstates. It is not possible to claim, in such
a non adiabatic protective measurement, that a time average is
being measured.

This appears to undermine the usual
Copenhagen probabilistic interpretation of the wave function
which
was developed on the basis of the usual measurement that results
in
a collapse of the wave function. Because probabilities
can be given physical meaning only by means of an
ensemble of identical systems. Whereas, protective
measurements,
which were unknown to the founders of quantum theory, enables
the
wave function of a single system to be determined up to phase.
The problems  associated with the usual measurement may be
indicative of new physics needed for its understanding. This new
physics may then negate the meaning given to the wave function
on
the basis of our present understanding or lack of understanding of
the usual measurement. It may therefore be safer to interpret the
wave function by means of the  protective measurement for
which
the physics is well understood.
\vglue 0.6cm
\line{\elevenbf ACKNOWLEDGEMENTS \hfil}
\vglue 0.4cm

I thank Y. Aharonov and H. R. Brown for useful discussions. This
work was
partially
supported by
NSF grant PHY-9307708 and ONR grant R\&T 3124141.
\vglue 0.6cm
\line{\elevenbf REFERENCES \hfil}
\vglue 0.4cm

\item{1.} P. A. M. Dirac, {\it The Principles of Quantum Mechanics}
(Oxford Univ. Press 1930); J. von Neumann, {\it Mathematical
Foundations of Quantum Mechanics} (Springer, Berlin 1932;
Princeton Univ. Press, Princeton 1955).
\item{2.} Y. Aharonov and L. Vaidman, Phys. Lett. A {\bf 178,} 38
(1993).
\item{3.}Y. Aharonov, J. Anandan, and L. Vaidman, Phys. Rev. A
{\bf
47,} 4616 (1993).
\item{4.} J. Anandan, Foundations of Physics
Letters {\bf 6,} 503 (1993).
\item{5.} This proposed experiment was realized during
discussions I had with R. Bhandari.
\item{6.} Y. Aharonov and J. Anandan, Phys. Rev. Lett. {\bf 58,}
1593 (1987).
\item {7.} M. V. Berry, Proc. Roy. Soc. A {\bf 392,} 45 (1984).
\item {8.} J. Anandan, Nature {\bf 360,} 307 (1992).
\item {9.} J. Anandan, Annales de l'Institut Henri Poincare,
{\bf 49,} 271 (1988).
\item{10.} J. Anandan, Phys. Lett. A {\bf 147,} 3 (1990).
\item{11.} J. Anandan, Found. of Physics {\bf 21,} 1265 (1991).
\item{12.} T. W. B. Kibble, Commun. Math. Phys. {\bf 65,} 189
(1979).
\item{13.} For a more mathematical treatment of the symplectic
structure see
refs. 9-12, and L. P. Hughston in {\it Twistor Theory}, ed. S. A.
Huggett (Marcel Dekker, 1995). The present work was was done
independently of
the latter paper, which contains interesting results on the
implications of the projective geometry of $\cal P$ to quantum
theory.
\item{14.} D. N. Page, Phys. Rev. A {\bf 36,} 3479 (1987).
\item{15.} J. Anandan and Y. Aharonov, Phys. Rev. Lett. {\bf 65,}
1697 (1990).
\item{16.} See for example R. Penrose in {\it Quantum Coherence
and Reality,} editors J. S. Anandan and J. L. Safko (World Scientific,
1994).
\item{17.}  C. Rovelli, Phys. Rev. {\bf A 50}, 2788 (1994).
\item{18.}J. Samuel and R. Nityananda, e-board: gr-
qc/9404051.\hfill
\item{19.} W. G. Unruh, Phys. Rev. {\bf A 50}, 883 (1994).
\item{20.} P. Ghose and D. Home, preprint (1993).
\item{21.} For a description of Einstein's hole arument and his
resolution of it, see {\it Relativity and Geometry} by R. Torretti
(Pergamon Press, Oxford 1983), 5.6.
\bye